# Deep graph embedding for prioritizing synergistic anticancer drug combinations

Peiran Jiang[a,b,#], Shujun Huang[c,#], Zhenyuan Fu[d], Zexuan Sun[a,e], Ted M. Lakowski[c], Pingzhao Hu[a,f*]

[a]Department of Biochemistry and Medical Genetics, University of Manitoba, Winnipeg, Manitoba, R3E 0J9, Canada;

[b]Department of Bioinformatics & Systems Biology, Huazhong University of Science and Technology, Wuhan, 430074, China;

[c]College of Pharmacy, University of Manitoba, Winnipeg, Manitoba, R3E 0T5, Canada;

[d]Tongji Medical College, Huazhong University of Science and Technology, Wuhan, 430030, China

[e]School of Mathematics and Statistic, Wuhan University, Wuhan, 430072, China

[f]Research Institute in Oncology and Hematology, CancerCare Manitoba, Winnipeg, R3E 0V9, Canada, Canada

[#]These authors equally contributed to this work.

[*]**Corresponding author at**: Department of Biochemistry and Medical Genetics, Room 308 - Basic Medical Sciences Building, 745 Bannatyne Avenue, University of Manitoba, Winnipeg, Manitoba, R3E 0J9, Canada.
Tel.: +1-204-789-3229 (P. Hu).
E-mail address: pingzhao.hu@umanitoba.ca (P. Hu).

Peiran Jiang: peiran@hust.edu.cn
Shujun Huang: huangs12@myumanitoba.ca
Zhenyuan Fu: u201612472@hust.edu.cn
Zexuan Sun: zexuansun@outlook.com
Ted M. Lakowski: ted.lakowski@umanitoba.ca
Pingzhao Hu: pingzhao.hu@umanitoba.ca

**Abstract**

Drug combinations are frequently used for the treatment of cancer patients in order to increase efficacy, decrease adverse side effects, or overcome drug resistance. Given the enormous number of drug combinations, it is cost- and time-consuming to screen all possible drug pairs experimentally. Currently, it has not been fully explored to integrate multiple networks to predict synergistic drug combinations using recently developed deep learning technologies. In this study, we proposed a Graph Convolutional Network (GCN) model to predict synergistic drug combinations in particular cancer cell lines. Specifically, the GCN method used a convolutional neural network model to do heterogeneous graph embedding, and thus solved a link prediction task. The graph in this study was a multimodal graph, which was constructed by integrating the drug-drug combination, drug-protein interaction, and protein-protein interaction networks. We found that the GCN model was able to correctly predict cell line-specific synergistic drug combinations from a large heterogonous network. The majority (30) of the 39 cell line-specific models show an area under the receiver operational characteristic curve (AUC) larger than 0.80, resulting in a mean AUC of 0.84. Moreover, we conducted an in-depth literature survey to investigate the top predicted drug combinations in specific cancer cell lines and found that many of them have been found to show synergistic antitumor activity against the same or other cancers *in vitro* or *in vivo*. Taken together, the results indicate that our study provides a promising way to better predict and optimize synergistic drug pairs *in silico*.

# 1. Introduction

Drug combinations, also known as combinatorial therapies, are frequently prescribed to treat patients with complex diseases especially cancers [1,2], which have been driven by many mechanisms concurrently [3]. The rationale of drug combination is that targeting multiple molecular mechanisms in cancer cells simultaneously can typically increase the potency of the treatment [1,2]. Thus, compared to monotherapies (i.e., single drug treatments), whose effectiveness may be limited, drug combinations have been reported with the potential to increase efficacy [1,2], decrease adverse side effects [4], or overcome drug resistance in cancer treatment [5]. However, a concurrent use of multiple drugs may sometimes cause adverse effects [6]. For example, the addition of panitumumab to bevacizumab and oxaliplatin- or irinotecan-based chemotherapy has been shown to lead to an increased toxicity and decreased progression free survival of metastatic colorectal cancer patients [6]. Therefore, it is critical to evaluate the effects of drug combinations in cancer cells and thereby identify those showing synergistic effects in a particular cancer type. Synergistic drug combinations exhibit greater total effect of the drugs than the additive effects of the individual drugs [7].

One of the challenges in studying drug synergy is that the possible number of drug combinations grows exponentially with the number of drugs under consideration, further expanded by the number of cancer types and drug dosages. Conventionally, effective drug combinations were proposed according to clinical trials, which are time- and cost-consuming, and what was worse, may expose patients to unnecessary or even harmful treatments [8,9]. More recently, high throughput screening (HTS) approaches have been extensively used to determine and evaluate effective combination strategies in a preclinical setting, which test an enormous number of drug combinations prescribed with different dosages and applied to different cancer cell lines [4,10,11]. An example is the study performed by O'Neil and collogues [4]. This study carried out 23,062 experiments on 583 drug combinations across 39 cell lines from various cancer types, recapitulating in vivo response profiles. Although they don't perfectly represent the original tumor tissues, cancer cell lines can be used to provide an alternative way for assessing the synergistic properties across drugs. Thus, data generated by HTS strategies enables the possibility of silico prediction of novel synergistic drug pairs, which can further guide *in vitro* and *in vivo* discovery of rational combination therapies.

A number of computational methods have been developed to predict anti-cancer drug synergy using chemical information from drugs, or molecular data from cancer cell lines, or both. The approaches rang from traditional machine learning models to deep learning methods. Sidorov and collogues utilized two machine learning methods (random forest (RF) and extreme gradient boosting (XGBoost)) to develop models for drug synergy prediction [12]. The models took the physicochemical properties of drugs as input and were trained on a per-cell line basis, which means each method (RF or XGBoost) was used to generate a model for each cell line. The XGBoost method demonstrated a slightly better prediction performance than the RF technique when they were evaluated in a new data set. As shown in [7], given a drug pair comprising drugs A and B and a particular cell line C, a deep learning-based regression model (termed DeepSynergy) was developed using both the chemical descriptors for drugs A and drug B and the gene expression profiles of the cell line C to predict the synergy scores of specific drug combinations on a given cell line. DeepSynergy demonstrated an improvement of 7.2% in its performance over Gradient Boosting Machines for drug synergy prediction task. Zhang and collogues [13] also proposed a deep learning-based model named AuDNNsynergy by integrating multi-omics data (i.e., the gene expression, copy number and genetic mutation data) from cancer cell lines to predict synergistic drug combinations. AuDNNsynergy outperformed the other four approaches, namely

DeepSynergy, gradient boosting machines, random forests, and elastic nets. Other studies, such as Hsu et al. [14], explored gene set-based approaches to predict the synergy of drug pairs. However, there are limited works applying the recently developed graph convolutional network (GCN) approaches [15] to predict drug synergy in cancers by integrating multiple biological networks. This study tried to develop GCN models to predict synergistic drug combinations in cancer cell lines by performing heterogeneous graph embedding from an integrated drug-drug combination, drug-protein interaction, and protein-protein interaction network.

## 2. Material and Methods

### 2.1. Data collection

Our study design is depicted in **Fig.1**. The GCN model for synergistic drug combination prediction was cell line-specific and based on three different types of subnetworks: drug-drug synergy (DDS) network, drug-target interaction (DTI) network, and protein-protein interaction (PPI) network. Data from various sources such as online databases and the published literature were collected to build the three networks (**Table 1**). Firstly, we obtained the DDS data from O'Neil *et al.*'s study [4]. This study contains 23,062 drug-drug combinations with the corresponding Loewe synergy scores tested across 38 drugs in 39 cell lines derived from 6 human cancer types. The Loewe synergy score is also termed as Loewe Additivity [16]. Secondly, the DTI data were acquired from the STITCH database [17], which provides voluminous interactions between chemical compounds and target proteins. More than 500,000 compounds and 8,900 proteins consist of 8,083,600 interactions. Of these DTIs, we included those that have been validated experimentally. Thirdly, we collected the PPI data from two comprehensive open access repositories, the BioGRID database [18] and the STRING database [19]. We further enriched these two databases with two additional PPI networks published by Menche *et al.* [20] and Rolland *et*

*al.* [21], resulting in a total of 719,402 experimentally validated physical interactions over 19,085 unique proteins.

The three types of subnetworks were used to construct a heterogenous network interactively. This final heterogeneous network is the intersection of heterogenous entities (i.e., proteins and drugs), and has their links from the three subnetworks [15]. Cell line specific networks were constructed by focusing on the links in a given cell line (**Fig.2**). In addition, only related PPIs and DTIs were preserved in a cell line-specific network. As a result, 39 cell line specific heterogeneous networks were established.

Table 1 The data sources of three types of interactions

| Data sources | Number of links | Number of entities |
| --- | --- | --- |
| I(DDS) | 23,052 DDS | 38 drugs, 39 cell lines |
| II(DTIs) | 8,083,600 DTIs | 519,022 drugs, 8,934 proteins, |
| III (PPIs) | 719,402 PPIs | 19,085 proteins |

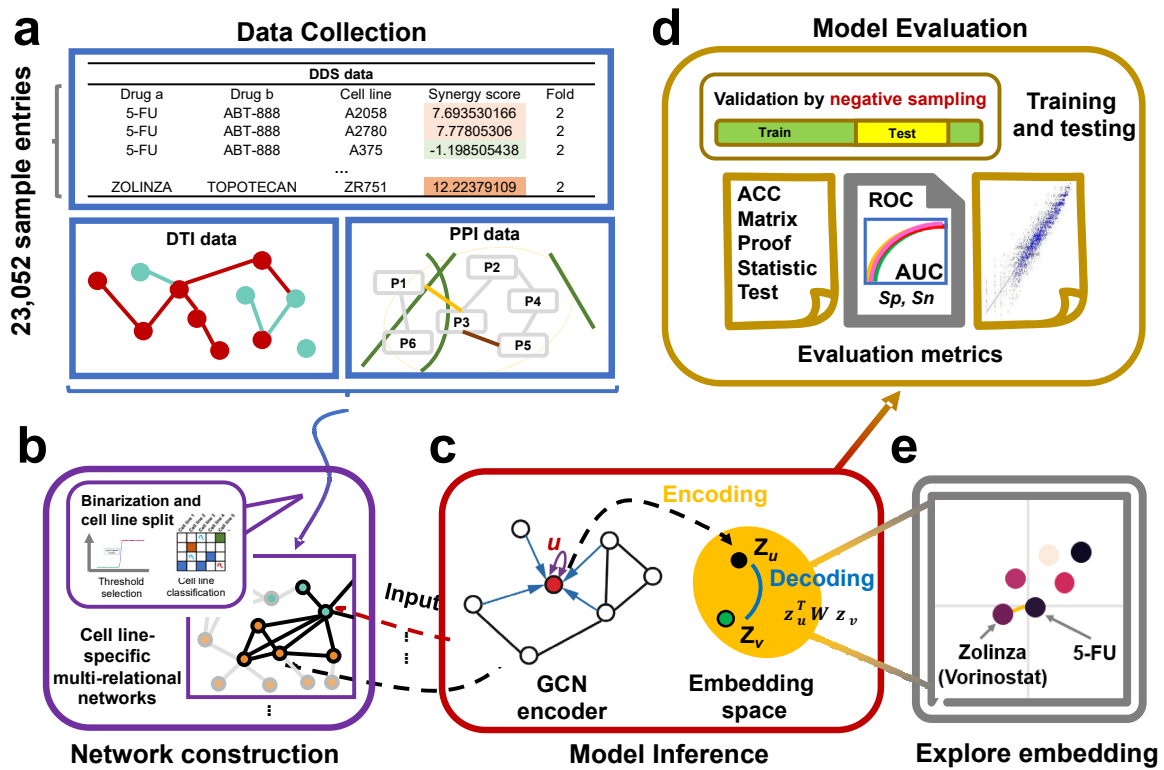

**Fig.1.** The study design. (a) Data collection. The drug-drug synergy (DDS) data, the drug-target interaction (DTI) data, and the protein-protein interaction (PPI) data were collected for the three subnetworks. (b) Network construction. For a given cell line, the synergy scores of drug pairs were binarized to construct the DDS subnetwork, which together with the DTI and PPI networks was further built the cell line-specific heterogenous network. (c) Model inference. The heterogenous network for a specific cell line is the input of the GCN encoder. Each encoded node is then mapped to an embedding space for representing the drug-drug synergy prediction in the new space. (d) Model evaluation. The negative sampling method together the accuracy, AUC, and Pearson correlation coefficient metrics were used. (e) Exploration of embedding space. t-SNE method was used to find the distribution of synergistic drug combinations.

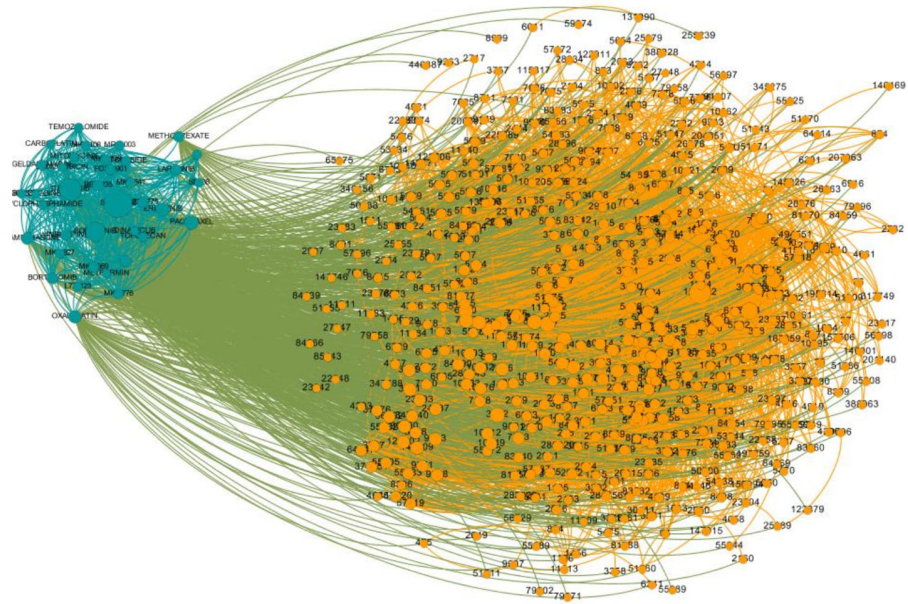

**Fig.2.** The cell line-specific heterogenous network derived from cell line CAOV3. The orange color represents the proteins (nodes) and their interactions (edges), the teal color represents the drugs (nodes) and their interactions (edges), and the olive color represents the interactions (edges) between the drugs and the proteins.

### 2.2. Graph Convolutional Network Encoder

As shown in [22], the prediction of synergistic drug combinations can be formulated as a link prediction problem using complex networks. In this study, we represented the different types of known links in the heterogeneous network belonging to each of the unique cell lines. Our research aim is to predict drug-drug synergic links using all the link information in the heterogeneous network [23]. The kind of prediction is related to semi-supervised learning in graphs such as GCN.

The GCN model is a neural network that operates on graphs and enables learning over graph structures. It is widely used as an encoder in different deep learning architectures. An encoder is a tool for mathematical transformation to map information from a space to another space (i.e., the embedding space). In this study, the encoder is based on GCN by operating the complex network to extract vital information in the graph.

To elucidate the GCN more clearly, the entities and their links in a network are usually represented by a defined graph $G = (V, E)$, where $V$ is a set of $N$ nodes such as drugs and proteins, and $E$ is a set of $M$ edges such as drug-drug links and drug-protein links. These $N$ nodes have numerical node feature vectors $x_1, x_2, x_3, ..., x_N \in \mathbb{R}^d$, where $d$ is the dimension of the feature vector. As for the edges, for example, $(v_i, v_j)$ represents the link between node $v_i$ and $v_j$. Although the convolution procedure in the Euclidean space is well-defined, it is not well-defined in the graph because of the uncertainty of topological structures in different nodes. In regular Euclidean arrays such as matrices and pixels in images, convolutional neural network (CNN) is of great efficiency. However, when considering graphs, the traditional CNN model is not as powerful as it in lattices. Many approaches were proposed to solve this problem in the past several years.

In 2016, Kipf *et al.* introduced a semi-supervised GCN model [24]. In this model, graph convolution operation on a graph was defined as a multi-layer propagation process. For the graph $G = (V, E)$, an adjacency matrix $A \in \mathbb{R}^{N \times N}$ and a degree matrix $D$ ($D_{ii} = \sum_j A_{ij}$) can be defined. The multi-layer model follows a layer-wise propagation rule as shown below:

$$H^{(l+1)} = \sigma\left(\widetilde{D}^{-\frac{1}{2}} \widetilde{A} \widetilde{D}^{-\frac{1}{2}} H^{(l)} W^{(l)}\right) \quad (1)$$

Here, $\widetilde{A} = A + I$ is the adjacency matrix $A$ of the undirected graph $G$ with added self-connections, $I$ is the identity matrix, $\widetilde{D}_{ii} = \sum_j \widetilde{A}_{ij}$, $W^{(l)}$ is a layer-specific weight matrix that is able to be trained, $\sigma(\cdot)$ is characterized as the activation function (i.e. ReLU$(\cdot) = \max(0, \cdot)$), $H^{(l)} \in \mathbb{R}^{N \times D}$ is the matrix of activations of the $l^{\text{th}}$ layer. Specially, $H^{(0)} = X$ ($X$ is the feature matrix consisting of $x_1, x_2, x_3, ..., x_N$). This form of propagation rule can be motivated and the final outcome after *k* layers of feature vector $x_i$ is the embedding vector $z_i$. A general GCN model can be considered as a function $g(X, A) = Z$ ($Z$ is the embedding feature matrix consisting of $z_1, z_2, z_3, ..., z_N$).

However, for the complex network-based drug-drug synergy prediction using the above mentioned GCN model, there is an obvious limitation. It considers only single node type and link type. This drawback restricts the usage of the GCN model in heterogenous networks. In 2018, Marinka *et al.* developed a multi-relational link prediction model called Decagon [15]. They applied this model to predict polypharmacy side effects and achieved state-of-the-art performance. In this study, we adopted the Decagon-based GCN algorithm, which is capable to extract information of different types of nodes and edges. For a given node in a graph, successive operations of graph convolutional layers integrate and transform information from its neighbors. In this architecture, the edges ($v_i, v_j$) from a given graph $G = (V, E)$ are divided into *r* types. Hence, the new representation of the edges is ($v_i, r, v_j$). For instance, 964 different relation types of drugs (side effects) were considered in Decagon. We continued to use this strategy for efficiently aggregating information from different edge types. There are 3 types of interactions ($r = 1, 2, 3$) in our drug-drug synergy prediction. The layer-wise propagation rule can be formulated as:

$$H^{(l+1)} = \sigma\left(\sum_r \widetilde{\widetilde{D}}^{-\frac{1}{2}} \widetilde{A} \widetilde{\widetilde{D}}^{-\frac{1}{2}} H^{(l)} W_r^{(l)}\right) \quad (2)$$

Here, $\widetilde{\widetilde{D}}$ is the adjusted Laplace matrix and $\widetilde{\widetilde{D}}_{ij} = N_r(v_i)N_r(v_j)$, where $N_r(v_i)$ is the number of the neighbors of node $v_i$ by link type *r*. This number is a constant for a given graph. $W_r^{(l)}$ is a trainable edge type-specific weight matrix of layer *l*. The same input and output forms as original GCN are maintained in this architecture. The first input layer could be numerical features, adjacency information or unique code of each node such as one-hot. Finally, the feature vector $x_i$ of a given node $v_i$ arrives in the ultimate embedding space as $z_i$.

## 2.3. Matrix bilinear decoder

After mapping each node into the embedding space, the primary task is to represent the drug-drug synergy in the new space. This process is defined as decoding. The goal of decoding is to reconstruct edge or node labels by the integrated information from the embedding space. Voluminous methods have been introduced to decode the embeddings. Matrix factorization is one of the easiest and most efficient way to perform the operation.

From the GCN encoder each node including either drugs or proteins is first encoded into a unique vector in the embedding space (**Fig.3**). Then we observe the embedding node set of all the drugs. Utilizing the embedding vectors $z_u$ and $z_v$, the synergy score between drug u and drug v is calculated by the following matrix in the bilinear form:

$$s(u, v) = z_u^T R_c z_v \qquad (3)$$

Where $R_c$ is a cell line-specific matrix to decode edges from node embedding vectors. $R_c$ is also trainable. The training process of $R_c$ is based on the cell line-specific heterogenous network. This is due to two reasons. Firstly, the synergistic effect is experimentally measured in different cell lines. Secondly, we expect to acquire the prediction among all the drug combinations across all the cell lines. $s(u, v)$ is the predicted synergy score of the combination of drug *u* and drug *v*. Not like the original Loewe synergy score, the predicted score is between 0 and 1. The higher value represents the larger potential of synergy. Using this approach, the matrix bilinear decoder is able to well comprehend the embedding space.

## 2.4. Model construction

To build the model for drug-drug synergy prediction, we first constructed the GCN deep encoder (**Fig.1c** and **Fig.3**). Our GCN encoder had an input layer and 4 hidden layers. Between each of two hidden layers, there was a ReLu activation function. The activation function of the last hidden layer was the sigmoid function:

$$f(y) = \frac{1}{1+e^{-y}} \quad (4)$$

Here, y is the output of previous layer and $f(y)$ is the embedding vector. 39 cell line specific-heterogeneous networks were input into the GCN encoder and the output was 39 cell line specific-embedding spaces including the embedding vectors for each drug.

Then from the embedding space, matrix decoder performed the mathematic operations to decode all given embedding vectors mentioned above. The matrix decoding was the 5$^{th}$ hidden layer. Finally, the result of the decoder was a cell line-specific synergy score matrix.

Optimization was implemented using the cross-entropy loss function:

$$L = -\frac{1}{N}\Sigma_i[y_i \log s_i + (1 - y_i) \log(1 - s_i)] \quad (5)$$

Here, $y_i$ is the real synergistic state of each drug combination, $s_i$ is the predicted synergy score for each combination, and $N$ is the number of drug combinations respectively. Backpropagation was carried out from the final loss back to each of the previous layers (**Fig.3**). We trained our full model including all the trainable parameters by this end-to-end method. It has been shown that the end-to-end learning can greatly improve the model performance because all the trainable parameters receive the gradients from the loss function jointly [25]. In this study, the loss propagates through both the GCN encoder and matrix decoder.

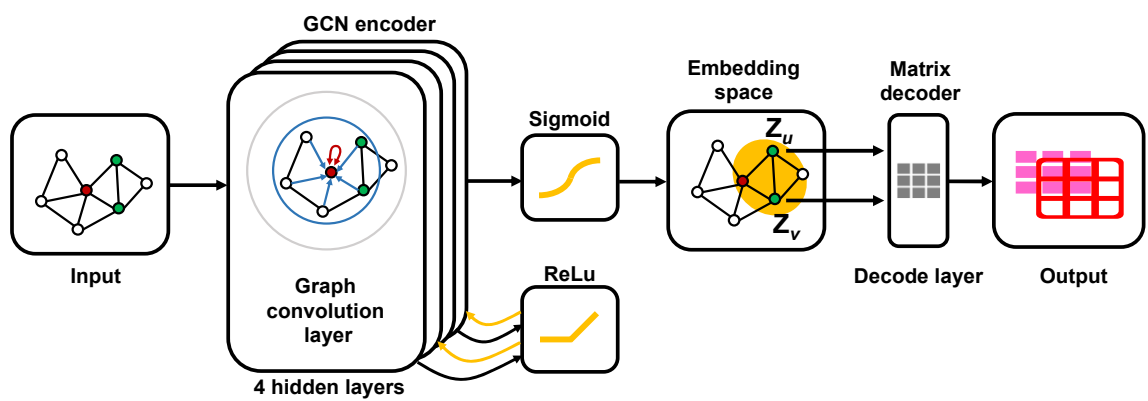

**Fig.3.** The workflow of GCN encoder and matrix decoder. There are 4 hidden layers in the GCN encoder. Between each of two hidden layers, there is a ReLu activation function. The output of the ReLu is the input for the next hidden layer. For the last hidden layer, we adopt a sigmoid activation function. The input of the GCN model is a graph and the output is an embedding vector for each node. Matrix decoder decodes the embedding vectors to predict the synergy score of any given drug combination.

**2.5 Model comparison**

To benchmark the performance of our method, we compared the GCN model to the other state-of-art machine learning and deep learning approaches, including support vector machine (SVM) [26], random forest [27], elastic net [28], and deep neural network (DNN) [29]. For the input features of these models, we utilized physiochemical properties of drugs including 1-D descriptors (i.e., molecular weight, molar refractivity, and logarithm of the octanol/water partition coefficient), 2-D descriptors (i.e., number of atoms, number of bonds, and connectivity indices), and fingerprints which represent unique molecular structure and properties in a particular complex form. We extracted these properties of each drug using the PaDEL software [30] with default settings. As a result, a total of 2,325 structural features were obtained to construct the feature vector for each drug. For a drug-drug pair, the two individual feature vectors were concatenated into a 4,650-demisional input feature vector.

In order to verify the power of graph structure in predicting drug-drug synergy, we also used the adjacency from the adjacency matrix as the features for DNN. For a given drug, the feature was the corresponding adjacency vector of the adjacency matrix in the cell-line specific heterogenous network. The true feature vector of a sample (drug-drug pair) was the concatenated vector of two drugs.

The DNN models maintained similar architectures as the GCN to avoid extra factors. There were 4 hidden layers in both the DNN using adjacency-specific features and the DNN using physiochemical-related features. The number of nodes in each of the 4 hidden layers was 1280, 640, 128, 48, respectively. For the activation function, ReLu was utilized between each of two hidden layers. The loss function was also cross-entropy loss.

### 2.6. Model evaluation

To evaluate the performance of our proposed GCN model, proper sampling methods were considered since k-fold cross validation in traditional machine leaning is not well compatible in complex network data. There are different sampling methods in graph-based problems. We used a recently developed method called negative sampling, which can precisely and robustly estimate the network-based model performance [31]. Specifically, in each cell line-specific heterogeneous network, we randomly sampled a fixed proportion (i.e., 10%) of drug-drug links [15]. Then random selection of edges was performed by randomly selecting nodes according to a sampling distribution [31,32]. In our model, 10% was set as the sampling rate.

To compare the model performance of SVM, random forest, elastic net, DNN (physiochemical) and DNN (adjacency), which were based on matrix-like feature vectors, we carried out 10-fold cross-validation strategy [33] to obtain unbiased performance.

Performance was evaluated individually in each cell line. Since we split the drug-drug links into training and testing edges by the negative sampling strategy, we utilized two performance metrics, that is, accuracy and area under the curve (AUC) of a receiver operating characteristic (ROC) curve, to measure the performance of our model. Accuracy is the proportion of drug-drug pairs that were correctly predicted. AUC indicates the general performance under different true positive rates (TPR, sensitivity) and false positive rates (FPR, 1-specificity). AUC is a robust metric at different discrimination cutoffs assigned to a given drug-drug pair for a synergy. AUC is a value between 0 and 1 and the larger the value the better the performance of the model. The synergy score was calculated over all the cell lines to obtain an averaged performance.

**2.7. Software and global parameters**

For deep learning-based methods (GCN and DNN), our software environment incorporated the Keras version 2.2.4 (with tensorflow 1.13.1 backend) from http://github.com/fchollet/keras, which is a high-level neural networks API, written in Python and capable of running fast parallel computing. For all machine learning-based approaches (SVM, random forest and elastic nets), we implemented scikit-learn version 0.21.2 (A powerful open source machined learning package in python) to achieve the performance.

We set mini batch size as 256 to ensure fast training and high accuracy. 20% dropout rate and 200 epochs of Adam optimizer [34] were used to avoid overfitting. The learning rate was set to 0.0001. Other training parameters including degree of momentum, strength of parameter regularization and initial weights were updated and optimized at the same time to reach optimal performance.

# 3. Results

## 3.1. Selection of the measured drug synergy score threshold

To construct the cell line-specific heterogeneous networks, we first selected the optimized synergy score threshold through the negative sampling strategy. The performance was analyzed for a series of synergy score thresholds from 0 to 60. This pre-predictive evaluation was performed by the averaged AUC across all the cell lines. Basic GCN model was utilized to train the model. As shown in **Fig.4**, the best performance was obtained when the synergy score threshold was set at 30.

Using the synergy score threshold of 30, we sequentially kept the drug-drug pairs with the scores equal or larger than 30 in the heterogeneous network. Cell line-specific networks were then constructed by using the links in a given cell line. 2,081 drug-drug pairs from 39 cell lines and six distinct tissue types were included.

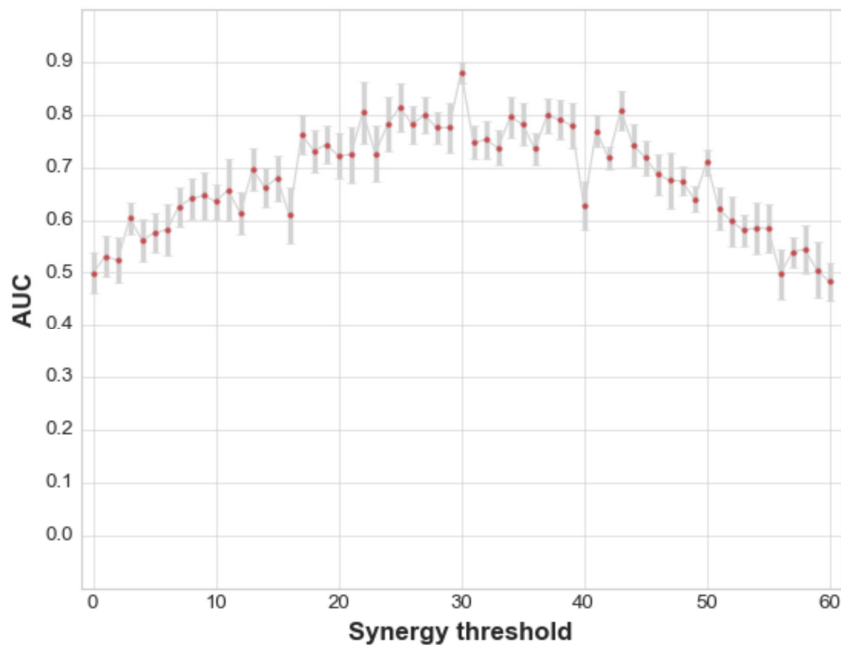

**Fig. 4.** The performance of different synergy thresholds. The performance measured by AUC was analyzed for synergy scores from 0 to 60 and the best threshold was obtained at 30

**3.2. Cell line-specific drug synergy prediction**

The trained model was used to predict the synergy scores of all drug combinations included in the network. The predicted values range from 0 to 1 and can be treated as probabilities to measure the synergy status of the drug combinations. According to the definition of Loewe synergy score, any score greater than 0 indicates the synergistic effect between the two drugs. In our predictions, this critical value was set to 0.5 since drug combinations with higher synergy score are of high value in experimental validation and clinical research.

In order to assess the performance of our model, we calculated the sensitivity and specificity for each cell line at different probability thresholds. The ROC curves were drawn and the AUC was obtained for each cell line (**Fig.5**). The average AUC across cell lines is 0.84, with COAV3 showing the largest AUC of 0.93 while COLO320DM showing the smallest AUC of 0.61. The accuracy for all cell line-specific GCNs were also calculated (**Fig.5**), resulting in a mean accuracy of 0.90. COLO320DM has the highest accuracy of 0.96 while MSTO has the lowest accuracy of 0.83. In general, the accuracy and AUC of the GCN models for most cell lines are stable and the majority of these models have high accuracy and AUC (e.g., > 0.80), suggesting the overall performance of our model is excellent. Furthermore, in a small proportion of the cell lines, the performance is perfect. This varied prediction performance indicates that the drug-drug synergy is highly dependent on the characteristics of cell lines and tissues.

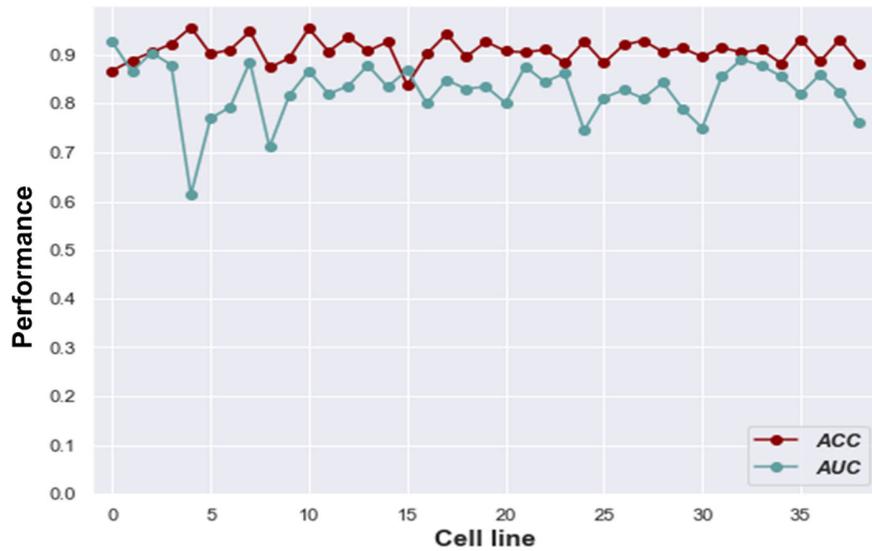

**Fig.5**. The performance of DDS prediction for all cell lines. Both ACC (accuracy) and AUC are shown in the same coordinate system. The x-axis is the cell line index. The y-axis is numeric from 0 to 1.

### 3.3. Investigation of prediction performance among tissues and drugs

To further understand the varied performance, we evaluated the correlation of the observed synergy scores and the predicted synergy scores at tissue and drug levels, respectively. We utilized Pearson correlation coefficient, a powerful and understandable method to check the consistency between the two variables, to further investigate the variability.

By integrating the predicted synergy scores with the measured synergy scores in the 39 cell lines from six tissues, we calculated the Pearson correlation (**Fig.6a**). The median of the coefficients is 0.64 for melanoma, 0.83 for ovarian, 0.67 for lung, 0.59 for colon, 0.71 for breast, and 0.68 for prostate. Among all the tissues, cell lines from ovarian show the highest median whereas those from prostate show the most concentrated distribution. The relatively high correlation suggests our model's consistency in tissue-wise aspects.

We further investigated the drug-wise correlation between predicted and measured synergy scores. For each drug, the correlation coefficient was averaged across all cell lines and existing drug combinations. The coefficients of drugs ranged from 0.39 to 0.90 (**Fig.6b**). For example, dasatinib used for treatment of chronic myeloid leukemia [35] has the highest correlation coefficient of 0.89. MK-2206, zolinza and MK-8669 also exhibit high correlation in the drug-specific prediction. The drug-wise analysis suggests that complicated pharmacological actions contribute to the variability of drug-drug synergy prediction.

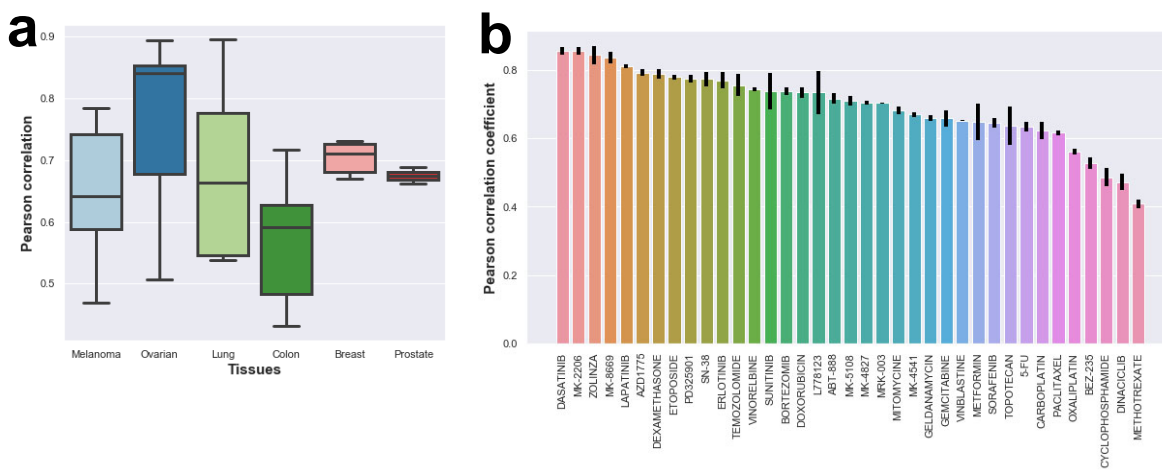

**Fig.6.** The Pearson correlation coefficients of the GCN models. (a) The boxplot shows the Pearson correlation coefficients between true and predicted synergy scores per tissue types. On the x-axis the tissue names are displayed. (b) The bar plot shows the Pearson correlation coefficients between true and predicted synergy scores per drugs. On the x-axis the drug names are displayed. The error bar is shown for each drug.

### 3.4 Data visualization and regression analysis

To better understand the consistency of the drug combinations in different cell lines at the same time, we constructed the 3-D matrix to illustrate the synergy distribution. Both experimental (blue dots) and predicted (orange dots) data were shown together in **Fig.7a**. Generally speaking, data from the two measurements indicate the similar patterns. Since the predicted scores covered both training and testing data, the 3-D synergy score distribution significantly supports that our model was well-trained and reached relatively high accuracy in the testing data.

Furthermore, in order to eliminate the potential bias in regression and illustrate the synergy score distribution more properly, we scaled the experimentally observed Loewe synergy scores to a fixed range of 0 to 1 using the min-max scaling method. We regressed the predicted synergy scores on the normalized measured synergy scores, which is shown in **Fig.7b** with an R-squared value of 0.768 ($p$-value < 0.05). This demonstrates that the predicted synergy score is highly consistent with the measured ones.

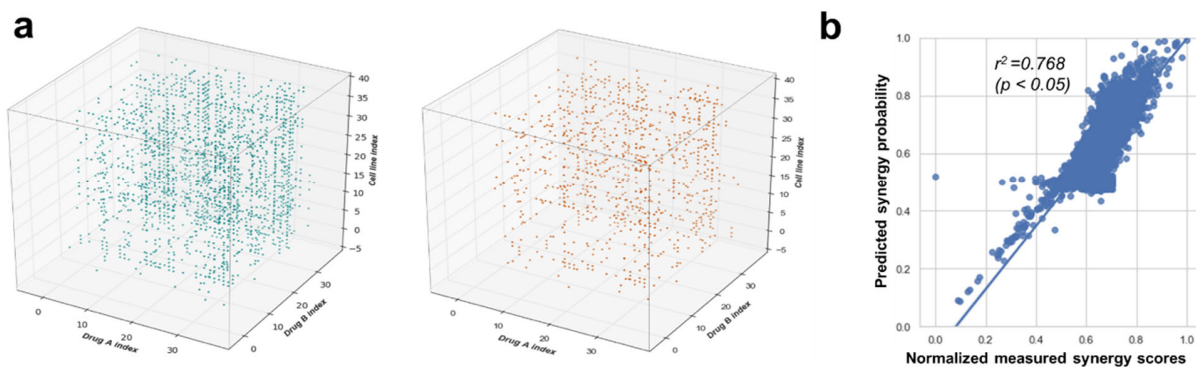

**Fig. 7.** The diagram of visualization and regression. (a) The 3-D matrix representation for experimentally measured drug synergy scores and predicted drug synergy scores. Each dot represents an experimental (blue, cutoff = 60, more than 60) or a predicted (orange, cutoff = 0.75) measurement of the synergy effect of drugs A and B in a specific cell line. The x axis is first drug index. The y axis is the second drug index. The z axis is the cell line index. (b) The regression of the predicted and measured synergy scores for all cell lines. Dots here are also flattened dots from the two 3-D matrices. The x-axis is the normalized measured synergy scores and y-axis is the predicted synergy probability (from 0 to 1).

### 3.5. Model comparison

We compared our GCN model with other four baseline methods. The mean and standard deviation (SD) of accuracies and AUCs of all cell lines for each method are listed in **Table 2**. The standard deviation can indicate the models' robustness. Our proposed GCN approach achieved the best performance in terms of both accuracy and AUC. The GCN method demonstrated an improvement of 11% in its AUC compared to the second-best method, which is the adjacency-based DNN. We believe this is because our GCN model successfully utilized the interconnections between heterogenous nodes. Among the four baseline methods, the DNN approach using either the adjacency features or the physiochemical features also showed a decent result. The DNN model with the physiochemical features was slightly better than that with the adjacency features in terms of AUC. Additionally, we also found that the performance of deep learning-based methods is better than that of other relatively traditional machine learning-based methods.

**Table 2** Performance of different classification methods. The reported Accuracy of AUC are the average of those for all cell lines.

| Method | Accuracy | | AUC | |
| --- | --- | --- | --- | --- |
| | mean | SD | mean | SD |
| GCN | 0.91 | 0.02 | 0.84 | 0.04 |
| DNN (adjacency) | 0.88 | 0.03 | 0.75 | 0.05 |
| DNN (physiochemical) | 0.83 | 0.05 | 0.81 | 0.02 |
| SVMs | 0.87 | 0.06 | 0.76 | 0.07 |
| Elastic Nets | 0.88 | 0.05 | 0.74 | 0.06 |
| Random Forests | 0.87 | 0.04 | 0.78 | 0.05 |

## 3.6. De novo prediction of drug synergy for particular cell lines

The drug pairs with the highest predicted probability for synergy were selected for each cell line. Of these 39 drug pairs, the BEZ-235/MK-2206 combination and the oxaliplatin/sunitinib combination ranked highest for the colon cancer cell line COLO320DM and the lung cancer cell line NCIH520, respectively. However, both of the two pairs show a low predicted probability of 0.1, and were thus removed. The remaining cell line-specific top predictions are listed in **Table 3**. To further examine the reliability of these top predicted drug combinations, we performed an in-depth literature survey and found that many of these pairs have been reported to show synergistic effects in cancer treatment. For instance, bortezomib and dasatinib have been used as lung cancer therapy recently (i.e. small cell lung cancer and non-small cell lung cancer) according to the studies [36,37]. Although some literatures reported that there was lung toxicity when using these two drugs [6,38], our model predicted them to have synergistic effect in lung cancer cell lines. Evidence from both cell line-level experiments [39] and clinic trials [40] indicate bortezomib has synergistic effect with dasatinib by inhibiting cell viability and promoting apoptosis within dasatinib-treated cells. This evidence is an example that our *de novo* predictions of drug-drug synergy are strongly supported by previous experiments.

Table 3 Top predicted synergistic drug combinations for each of the 39 cancer cell lines. Colon cancer cell line COLO320DM and the lung cancer cell line NCIH520 were not included in the table due to the low predicted probability synergy score of the top drug combinations in the two cell lines.

| Cell line | Cancer | Drug A | Drug B | Probability for synergy |
|---|---|---|---|---|
| OCUBM | Breast | ABT-888 | MK-8669 | 0.98 |
| ZR751 | Breast | AZD1775 | BEZ-235 | 0.92 |
| MDAMB436 | Breast | BEZ-235 | Temozolomide | 0.86 |
| T47D | Breast | Sunitinib | BEZ-235 | 0.86 |
| KPL1 | Breast | MK-8669 | MK-2206 | 0.82 |
| EFM192B | Breast | Dasatinib | MK-8669 | 0.78 |
| HT29 | Colon | MK-4827 | Temozolomide | 0.95 |
| RKO | Colon | MK-2206 | MK-8669 | 0.88 |
| SW620 | Colon | Dasatinib | Sunitinib | 0.87 |
| SW837 | Colon | Lapatinib | MK-2206 | 0.87 |
| HCT116 | Colon | BEZ-235 | MK-8776 | 0.82 |
| LOVO | Colon | Lapatinib | Dasatinib | 0.82 |
| DLD1 | Colon | Sunitinib | Temozolomide | 0.73 |
| SKMES1 | Lung | MK-4827 | SN-38 | 0.93 |
| NCIH460 | Lung | BEZ-235 | MK-4827 | 0.90 |
| MSTO | Lung | Bortezomib | Dasatinib | 0.87 |
| NCIH23 | Lung | Temozolomide | MK-4827 | 0.84 |
| A427 | Lung | MK-8669 | Temozolomide | 0.82 |
| NCIH1650 | Lung | Dasatinib | MK-8669 | 0.81 |
| NCIH2122 | Lung | MK-4827 | Temozolomide | 0.68 |
| NCIH520 | Lung | Oxaliplatin | Sunitinib | 0.10 |
| SKMEL30 | Melanoma | MK-8776 | MK-8669 | 0.98 |
| A375 | Melanoma | BEZ-235 | Temozolomide | 0.96 |
| UACC62 | Melanoma | MK-8669 | MK-4827 | 0.96 |
| A2058 | Melanoma | MK-8776 | Temozolomide | 0.89 |
| RPMI7951 | Melanoma | AZD1775 | MK-8669 | 0.84 |
| HT144 | Melanoma | BEZ-235 | MK-8669 | 0.62 |
| OV90 | Ovarian | Vinorelbine | MK-8776 | 0.97 |
| PA1 | Ovarian | BEZ-235 | MK-4827 | 0.94 |
| SKOV3 | Ovarian | MK-8669 | MK-4827 | 0.93 |
| UWB1289BRCA1 | Ovarian | BEZ-235 | Temozolomide | 0.91 |
| A2780 | Ovarian | MK-8669 | MK-2206 | 0.85 |
| CAOV3 | Ovarian | Etoposide | MK-2206 | 0.83 |
| OVCAR3 | Ovarian | Dasatinib | MK-8776 | 0.82 |
| UWB1289 | Ovarian | AZD1775 | BEZ-235 | 0.80 |
| ES2 | Ovarian | Sunitinib | BEZ-235 | 0.75 |
| VCAP | Prostate | BEZ-235 | MK-4541 | 0.93 |
| LNCAP | Prostate | BEZ-235 | Geldanamycin | 0.77 |

## 3.7. Exploration of embedding space of drug synergy prediction

From the decoder, we obtained the embedding vector for each drug based on the cell line specific heterogeneous network. The embedding space represents the predictions to some extent. Hence, we used a data dimension reduction method t-SNE to further explore the potential hidden patterns in this space [41].

The t-SNE method is able to cluster embedding vectors of drugs into a visible space such as 2D Euclidean space. We showed the result of t-SNE in a particular embedding space of cell line COLO320DM in **Fig.8**. It shows the drugs with higher probability of the synergistic effect, such as MK-8669/MK-2206 and sunitinib/dasatinib, were clustered together in the 2-D space. MK-8669 and MK-2206 share a short distance in the cluster and their synergistic effect has been reported by a previous study [42].

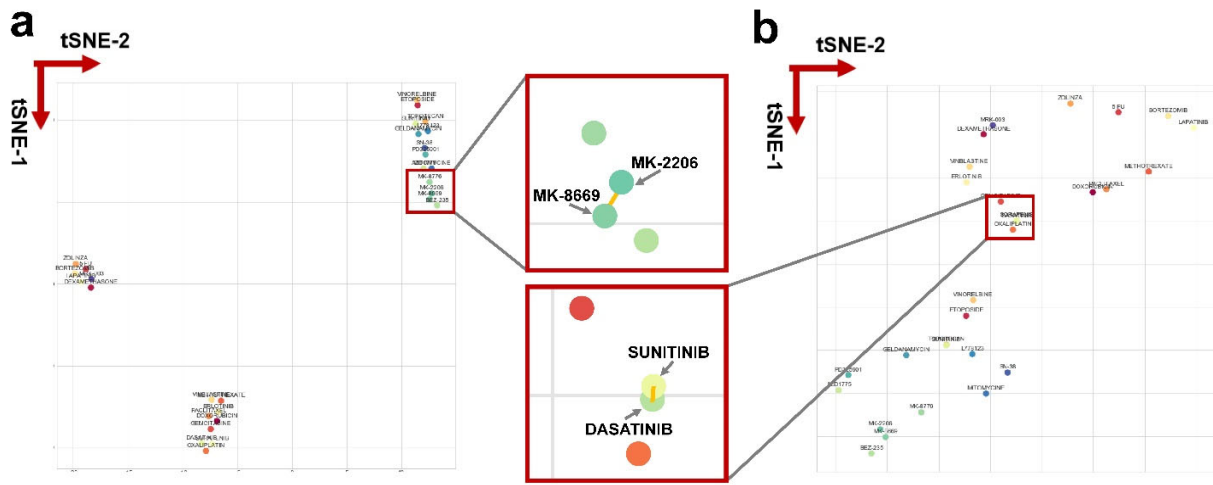

**Fig. 8.** Visualization of synergistic effects by t-SNE to explore the embedding space. The left panel (a) is the t-SNE result of the cell line KPL1-specific embedding space and the right panel (b) is the t-SNE result of the cell line SW620-specific embedding space. Two red frames in the middle are the magnifications in particular areas in (a) and (b). The x-axis is the first dimension of t-SNE and the y-axis is the second dimension of t-SNE. Each dot is a representation of a specific drug.

## 4. Discussion

Given the enormous number of drug combinations, experimentally screening all possible pairs is unfeasible in terms of cost and time. Thus, computational methods have been extensively used to predict potential synergistic drug combinations. In this study, we have developed a GCN-based model, which can predict synergy scores of drug combinations in particular cancer cell lines. Although it has been widely used in social network and knowledge graph prediction problems, GCNs have not until recently been introduced to the field of computational biology to predict sides effects caused by drug-drug interactions [15]. GCNs have not yet been used for prediction problems in drug synergy. Our GCN-based model for predicting synergistic drug combination was trained for each cell line and demonstrated a high accuracy, with a mean AUC of 0.84 (the minimum is 0.61, the maximum is 0.93) and a mean ACC of 0.91 (the minimum is 0.83, the maximum is 0.96), respectively. When treating the prediction task as a regression instead of a classification problem, the mean Pearson correlation coefficient between the measured and the predicted synergy scores of our GCN method for drug pairs in all cell lines was 0.70.

It is noteworthy that the GCN models from some cell lines performed better than others in terms of AUC. As an example, models for the CAOV3 (ovarian cancer) and A427 (lung cancer) cell lines are the two best-performing ones among the 39 cell line-specific GCN models, with an AUC larger than 0.90. The variability could be partly explained by the difference in the number of all conceivable drug combinations from each cell line. For example, some cell lines comprise ~700 tested drug combinations whereas others include approximate 500 screened drug combinations, leading to the varying training set size among cell lines. In addition, each cell line constitutes a different problem instance in this study. Even if training set size, features and classifiers are the same, the modeled relationship between drug synergy and features depends on training set composition and cell line properties due to the fact that the performance of supervised learning algorithms varies depending on the problem instance [43].

Among the top predicted drug pairs for the 39 cell lines (**Table 2**), many of them have been reported to be synergistic in the literature. For example, MK-8669 is a mTOR inhibitor [44] while MK-2206 is an Akt inhibitor [45]. For the estrogen receptor (ER)-positive breast cancer cell line KPL1 [46], the combination of these two agents shows the highest predicted synergy score, which is in accordance with a phase I clinical trial [42]. In this clinical study [42], a combination of MK-8669 and MK-2206, with the aim to completely block the PI3K/Akt/mTOR signaling pathway required for tumor growth, showed promising activity and good tolerability in ER-positive breast cancer patients with the PI3K/Akt/mTOR pathway addiction. The combination of MK-8669 and MK-2206 also shows the highest synergistic score predicted by both the RKO- and A2780-specific GCN models. RKO is a colon cancer cell line while A2780 is an ovarian cancer cell line. Since the PI3K/Akt/mTOR signaling pathway is aberrantly activated to sustain the growth and survival of tumor cells in many cancer types, including human breast, colon, and ovarian cancers [47], MK-8669 in combination with MK-2206 may act synergistically to block the PI3K/Akt/mTOR pathway in colon and ovarian cancer cells. For another triple negative breast cancer (TNBC) cell line MDAMB436 [48], BEZ-235 and temozolomide were predicted as the top synergistic pair. BEZ-235 is a novel dual PI3K and mTOR inhibitor and has been widely used in preclinical studies for various cancers including glioblastoma multiforme (GBM), breast, colorectal and lung cancers [49]. Temozolomide is a DNA alkylating agent and has been reported to induce cell apoptosis by the inhibition of mTOR signaling in GBM cells [50]. Compared with temozolomide or BEZ-235 monotherapy, a combination of the two drugs has been found to more effectively inhibit GBM cell proliferation, invasion, migration and induce apoptosis *in vitro* by repressing the PI3K/Akt/mTOR pathway singling activity [51]. The PI3K/Akt/mTOR signaling pathway is one of the most frequently altered pathways in TNBC [52]. Thus, the combination of temozolomide and BEZ-235 may be an effective treatment for TNBC, for which very limited targeted therapies exist currently. The combination of BEZ-235 and temozolomide was also predicted as the top synergistic pair for another two cell lines, the melanoma cell line A375 and the ovarian cancer cell line UWB1289BRCA1, for which constitutive PI3K/Akt/mTOR pathway activation has been observed [47]. Thus, our data suggests that BEZ-235 combined with temozolomide may have the potential to treat melanoma and ovarian cancer patients. Another combination of bortezomib and dasatinib has been predicted as synergistic pair in the lung cancer cell line MSTO by our GCN model and has experimentally shown synergistic antitumor activity in myeloma cell lines [53] and gastrointestinal stromal tumor cell lines [39]. Bortezomib is a small-molecule proteasome inhibitor and has activity in lung cancer both as a single drug and in combination with drugs commonly used in lung cancer [54]. Dasatinib is an inhibitor of Src family kinases and has modest clinical activity in lung cancer patients as a single drug in a phase II study [55]. Therefore, the combination of dasatinib and bortezomib may improve the treatment of lung cancer. The GCN model has also predicted some novel synergistic drug combinations, such as BEZ-235/geldanamycin in LNCAP (prostate cancer) and BEZ-235/MK-4541 in VCAP (prostate cancer). Geldanamycin is the first natural HSP90 inhibitor and has demonstrated antiproliferative and cytotoxic effects in both human prostate cancer cell lines [56] and prostate xenograft tumors [57]. The PI3K/mTOR dual inhibitor BEZ-235 combined with an HSP90 inhibitor (NVP-AUY922) has been reported to act synergistically to inhibit tumor cell proliferation and induce apoptosis in cholangiocarcinoma cell lines [58]. HSP90 is often overexpressed in prostate cancer cells, making it a potential therapeutic target for prostate cancer [59]. The PI3K/Akt/mTOR pathway also plays an important role in prostate cancer cell survival, apoptosis, metabolism, motility, and angiogenesis [47]. These findings suggest that the combination of BEZ-235/geldanamycin may exhibit synergistic antitumor activity against prostate cancer, for which further experimental validation is needed. MK-4541 is a novel selective androgen receptor modulator and found to exert anti-androgenic activity in the prostate cancer xenograft mouse model [60]. Androgens are critical for the development, growth, and maintenance of male sex organs and can drive prostate cancer initiation [60]. Therefore, the dual PI3K/mTOR inhibitor BEZ-235 might synergize with the androgen receptor inhibitor MK-4541 to treat prostate cancer.

Although a large publicly available synergy dataset has been used in this study [4], there is a limited number of different cell lines, drugs, and drug combinations. In this cell line-specific GCN method, some drugs have been tested in combination with only a few other drugs. This means these drugs do not have a good number of links with other drugs for the neural network model to be trained. Therefore, we removed these drugs in the cell line(s), which led to a decrease in the train set size. However, this limitation can be overcome as more HTS drug combination datasets are generated.

## 5. Conclusions

In this study, we utilized the graph convolutional network method to develop GCN models, which can predict synergistic drug combinations for 39 cell lines derived from six major cancer types, including breast, colon, lung, melanoma, ovarian, and prostate cancers. For the 39 cell line-specific GCN models we built, the mean AUC is 0.84 while the mean Pearson correlation coefficient between the measured and the predicted synergy scores is 0.70. Remarkably, we found that many synergistic combinations among our top predictions for a particular cancer type have been reported in the treatment of the same or other cancer types in the literature. Overall, given the prediction performance, the GCN models could be a valuable *in silico* tool for predicting novel synergistic drug combinations and thus guide *in vitro* and *in vivo* discovery of rational combination therapies.

**Abbreviation**

| | |
|---|---|
| GCN | Graph Convolutional Network |
| AUC | area under the curve |
| HTS | high throughput screening |
| RF | random forest |
| XGBoost | extreme gradient boosting |
| DDS | drug-drug synergy |
| DTI | drug-target interaction |
| PPT | protein-protein interaction |
| CNN | convolutional neural network |
| ROC | receiver operating characteristic |
| TPR | true positive rate |
| FPR | false positive rate |
| SVM | support vector machine |
| DNN | deep neural network |
| ACC | accuracy |
| SD | standard deviation |
| ER | estrogen receptor |
| TNBC | triple negative breast cancer |

GBM    glioblastoma multiforme

**Conflicts of Interest**

The authors declare no conflict of interest.

**Funding**

This research was supported in part by Canadian Breast Cancer Foundation, Natural Sciences and Engineering Research Council of Canada, Mitacs and University of Manitoba.